\documentclass[reprint,aps,pra,twocolumn,floats,superscriptaddress,]{revtex4-1}
\usepackage{bm}
\usepackage{graphicx}
\usepackage{times}
\usepackage{hhline}
\usepackage{array}
\usepackage{amsmath}

\newcolumntype{C}[1]{>{\centering\arraybackslash}m{#1}}
\usepackage{color}

\begin{document}

\author{W. Pacuski}\email{wojciech.pacuski@fuw.edu.pl}
\affiliation{Institute of Experimental Physics, Faculty of Physics, University of Warsaw, ul. Pasteura 5, 02-093 Warsaw Poland}

\author{J.-G. Rousset}
\affiliation{Institute of Experimental Physics, Faculty of Physics, University of Warsaw,
ul. Pasteura 5, 02-093 Warsaw Poland}
\affiliation{CNRS, Institut N\'{e}el, "Nanophysique et semiconducteurs"
group, F-38000 Grenoble, France}

\author{V. Delmonte}
\author{T. Jakubczyk}
\affiliation{CNRS, Institut N\'{e}el, "Nanophysique et semiconducteurs"
group, F-38000 Grenoble, France}
\affiliation{Univ. Grenoble Alpes, F-38000 Grenoble, France}

\author{{K.~Sobczak}}
\affiliation{Biological and Chemical Research Centre, University of Warsaw, \.{Z}wirki i Wigury 101, 02-089 Warsaw, Poland}

\author{{J.~Borysiuk}}
\author{K.~Sawicki}
\author{E.~Janik}
\affiliation{Institute of Experimental Physics, Faculty of Physics, University of Warsaw,
ul. Pasteura 5, 02-093 Warsaw Poland}

\author{J. Kasprzak}\email{jacek.kasprzak@neel.cnrs.fr}
\affiliation{CNRS, Institut N\'{e}el, "Nanophysique et semiconducteurs"
group, F-38000 Grenoble, France}
\affiliation{Univ. Grenoble Alpes, F-38000 Grenoble, France}

\title{Antireflective photonic structure for coherent nonlinear spectroscopy\\ of single magnetic quantum dots}

\begin{abstract}

This work presents epitaxial growth and optical spectroscopy of CdTe
quantum dots (QDs) in (Cd,Zn,Mg)Te barriers placed on the top of
(Cd,Zn,Mg)Te distributed Bragg reflector. The formed photonic mode
in our half-cavity structure permits to enhance the local excitation
intensity and extraction efficiency of the QD photoluminescence,
while suppressing the reflectance within the spectral range covering
the QD transitions. This allows to perform coherent, nonlinear,
resonant spectroscopy of \emph{individual} QDs. The coherence
dynamics of a charged exciton is measured via four-wave mixing, with
the estimated dephasing time $T_2=(210\,\pm\,40)$\,ps. The same
structure contains QDs doped with single Mn$^{2+}$ ions, as detected
in photoluminescence spectra. Our work therefore paves the way
toward investigating and controlling an exciton coherence coupled,
via $s$,$p$-$d$ exchange interaction, with an individual spin of a
magnetic dopant.

\end{abstract}

\maketitle

\newcommand{\Ea}{\ensuremath{{\cal E}_1}}
\newcommand{\Eb}{\ensuremath{{\cal E}_2}}
\newcommand{\Er}{\ensuremath{{\cal E}_{\rm R}}}

\section{Introduction}
\label{Introduction}

Distributed Bragg Reflectors (DBRs) are typically applied to enhance\cite{ShyhWang_IEEE_1974,Sadowski_PRB1997,Benisty_JQE98_01,Ng_APL_1999,Tawara00,guo00,Pac_APL2009,Gust_Nanotechnology_2009,Rou_JCG2013,Koba_JEWA2013,Koba_EPL2014,Jomek_ACSnano2014}
the reflectance ($R$),
but the opposite result --- a very low $R$ --- can be also designed
and obtained with the use of
DBRs\,\cite{Schubert_OE_2008,Chhajed_APL_2008}. Such antireflection
(AR) stacks are particularly useful in resonant optical
spectroscopy, where the reflected laser light dominates the
investigated signal substantially. In particular, using four-wave
mixing (FWM) for probing the coherence of single quantum dots (QDs),
necessitates efficient injection of the optical excitation into the
hetero-structure, so as to reach the field amplitude at the QD
location sufficient to induce its FWM. This condition has been
achieved by employing AR planar samples\,\cite{Langbein_OL_2006,Patton_PRB_2006,KasprzakNJP13} and, more recently, by suitably
designed nanophotonic devices\,\cite{KasprzakNatMater10,FrasNatPhot16,MermillodPRL16}. Here, we realize half-cavity AR planar structures and employ them to investigate FWM
of individual CdTe QDs. As the same structure contains QDs with
incorporated single Mn$^{2+}$ ions, the appealing prospect of our
work is to reveal the exciton's coherence coupled with a magnetic
dopant via $s$,$p$-$d$ exchange interaction.

\begin{figure*}[t]
 \centering
\includegraphics[width=1\textwidth]{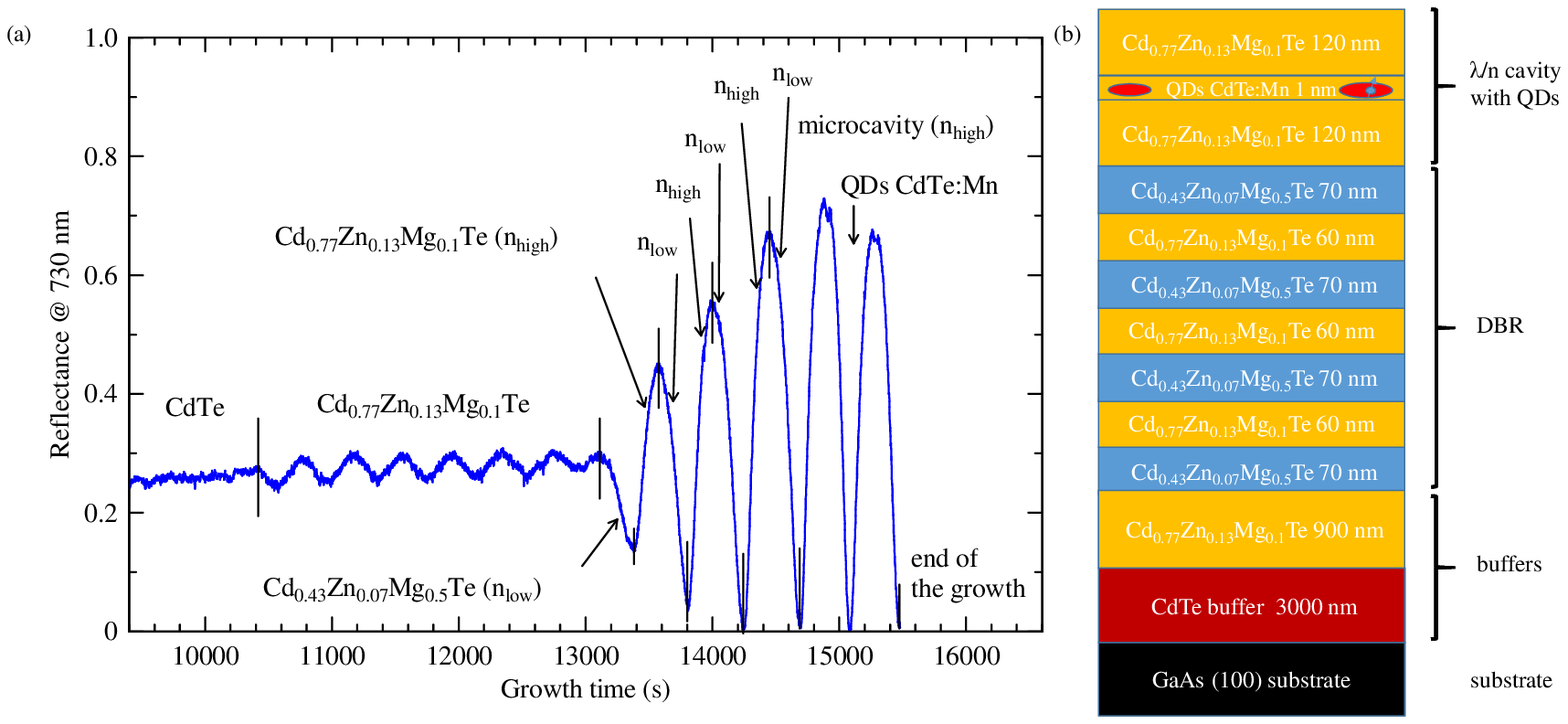}
\caption{(a) The reflectance time-trace at $\lambda = 730\,$nm
in-situ measured during the sample growth. (b) The sample layout.}
\label{fig:structure}
\end{figure*}

There are two basic approaches to AR structures on the top of high
refractive index ($n$) semiconductors. Firstly, one can deposit a
single layer with a lower refractive index equal to $\sqrt{n}$ and
thickness of $\lambda/(4n)$, where $\lambda$ is the targeted
wavelength for the lowest reflectance. While this method offers a
broad AR spectral range, its implementation is conditioned by the
availability of materials with a required value of $n$. Secondly,
one can use symmetric microcavities composed of two DBRs with the
nominally the same $R$. The cavity mode exhibits a pronounced dip in
the $R$ spectrum, which should be accurately fitted to the
wavelength of interest. In this work, we implement the AR strategy
based on a cavity system. In order to maximize the spectral width of
the cavity resonance, we have designed and grown a low Q-factor
structure ($Q\simeq20$): a half-cavity is formed by four bottom
Bragg pairs and the spacer containing the QD layer, also acting as a
semiconductor/air interface at the top.

\section{Growth and the sample layout}

The sample was grown using molecular beam epitaxy (MBE) model SVT
Associates. {Our molecular sources were standard Knudsen cells loaded with commercially available materials: Zn, Cd, and Te with 7~N purity; Mg and Mn with 6~N purity.
Chamber pressure was at level of about $2*10^{-9}$~Torr.} The growth was assisted by \textit{in-situ} optical
reflectance (model Filmetrics F-30) with a white light incident
beam perpendicular to the sample surface. The system is equipped
with a heated optical viewport, so that it was possible to perform
spectroscopic measurement during a long growth sequence, while
avoiding material deposition on the window. For the sample described
in this work, we monitored $R$ versus time at $\lambda = 730\,$nm,
as shown in Fig.\,\ref{fig:structure}\,(a). In our method, this
choice defined the resonant wavelength of the cavity for a given
spot of the sample (the substrate was not rotated during the
growth). However, other zones of the sample exhibit either thinner
layers with cavity resonance at shorter wavelength or thicker layers
with resonance at longer wavelength.

{
The growth was performed at constant temperature T = 345$^\circ$ C,} on a 2-inch GaAs:Si substrate oriented
(100) with 2$^{\circ}$ off. We started by depositing a 3$\,\mu$m
thick CdTe buffer. Since CdTe is absorbing at $\lambda=730\,$nm, the
corresponding oscillations of $R$ were rapidly suppressed and a
characteristic $R=0.25$ was measured before time $t = 10400\,$s. As
depicted in Fig.\,\ref{fig:structure}\,(b), we next grew a
Cd$_{0.77}$Zn$_{0.13}$Mg$_{0.10}$Te buffer lattice matched to
subsequent DBR layers. Due to the presence of Mg and Zn, such buffer
exhibits a slightly lower refractive index than CdTe and $250\,$meV
larger energy gap than CdTe. Therefore,
Cd$_{0.77}$Zn$_{0.13}$Mg$_{0.10}$Te buffer is transparent for
$\lambda = 730$\,nm, generating interferences in the $R(t)$ trace
(time between 10400\,s and 13100\,s in Fig.\,\ref{fig:structure}\,a)
during the growth of this material. {Observation of $R(t)$ interferences allowed us to determine growth rate v = 1.1 $\mu$m/h.
 Next, when $R(t)$ reached a local maximum,} we started the growth of
lattice matched DBRs composed of Cd$_{0.43}$Zn$_{0.07}$Mg$_{0.50}$Te
(low $n$ layer) and Cd$_{0.77}$Zn$_{0.13}$Mg$_{0.10}$Te (high $n$
layer)\,\cite{Rou_JCG2013,Rousset_APL2015,Mirek_PRB_2017}. Adjacent layers with
various composition of Mg were obtained using a pair of Mg sources,
such that either one or two sources were opened at a time. An exact
time-point for growing DBRs was not predefined in the programmed
sequence. Instead, it was adjusted manually by closing one of the Mg
shutters, as soon as the minimum of $R$ was reached, while keeping
opened both Mg shutters when the maximum of $R$ was attained (see
Fig.\,\ref{fig:structure}\,(a)). As expected, during the initial
stage of a DBR growth, after each pair, the contrast between the
maximum and minimum value of $R$ was increasing. Interestingly, $R$
reached the minimum value below 1$\%$ already after depositing the
fourth DBR pair. Since this is what required for an efficient AR
structure and further DBR growth would only increase reflectance,
we completed the growth of DBRs already after four pairs and we
completed $\lambda/n$ cavity composed of the high $n$ material
Cd$_{0.77}$Zn$_{0.13}$Mg$_{0.10}$Te, enclosing CdTe:Mn QDs in the
middle. As shown in Fig.\,\ref{fig:structure}\,(a), fabricating
thickness equal to $\lambda/(2n)$ or $\lambda/(n)$ of high index
material does not affect $R$ at $\lambda = 730$\,nm. Our half-cavity
structure is formed and terminated by the semiconductor/air
interface on the top.

QDs are formed out of about 1\,nm (3 monolayers) thick CdTe:Mn layer
grown at the same conditions as Cd$_{0.77}$Zn$_{0.13}$Mg$_{0.10}$Te
and the remaining top part of the sample.
{Basing on giant Zeeman effect\cite{GajDMSbook} measured for thicker CdTe:Mn layers grown in similar conditions, expected concentration of Mn in our CdTe:Mn layer is about $x=0.1\%$.} As a matter of fact, this layer forms a thin quantum well exhibiting interface fluctuations,
leading to formation of QDs. The growth of similar QDs, but in
Cd$_{0.7}$Mg$_{0.3}$Te barriers, was previously
reported\,\cite{Besombes_PRB2014}. Here, we used even lower barriers
and consequently we shifted the QD emission energy to spectral
region where the absorption of (Cd,Zn,Mg)Te is low enough to realize
microcavities.% (see Fig.\,\ref{fig:PL-ref}).

\begin{figure}
 \centering
\includegraphics[width=0.6\linewidth]{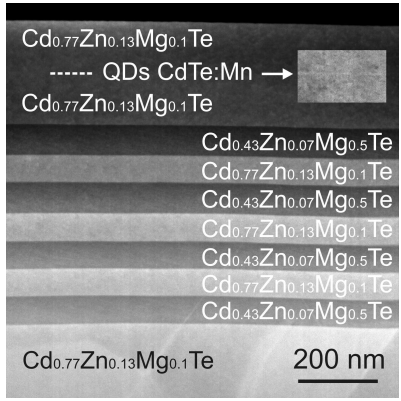}
\caption{{Image of the studied structure. DBR layers - alternating bright and dark layers are well resolved. The inset in the right corner is a magnified part of the cavity area containing CdTe QDs. The image is prepared using high angle annular dark field (HAADF) scanning transmission electron microscopy (STEM).}}
\label{fig:TEM}
\end{figure}

{Microscopy image of the most important parts of
studied structure is shown in Fig.~\ref{fig:TEM}. DBR layers, seen
as alternating bright and dark layers, are well resolved. Also, the QD layer is resolved in the middle of the cavity, it is seen as a thin
trace in the magnified part of the cavity.}

\section{Optical spectroscopy}

\begin{figure}[t]
 \centering
\includegraphics[width=1\columnwidth]{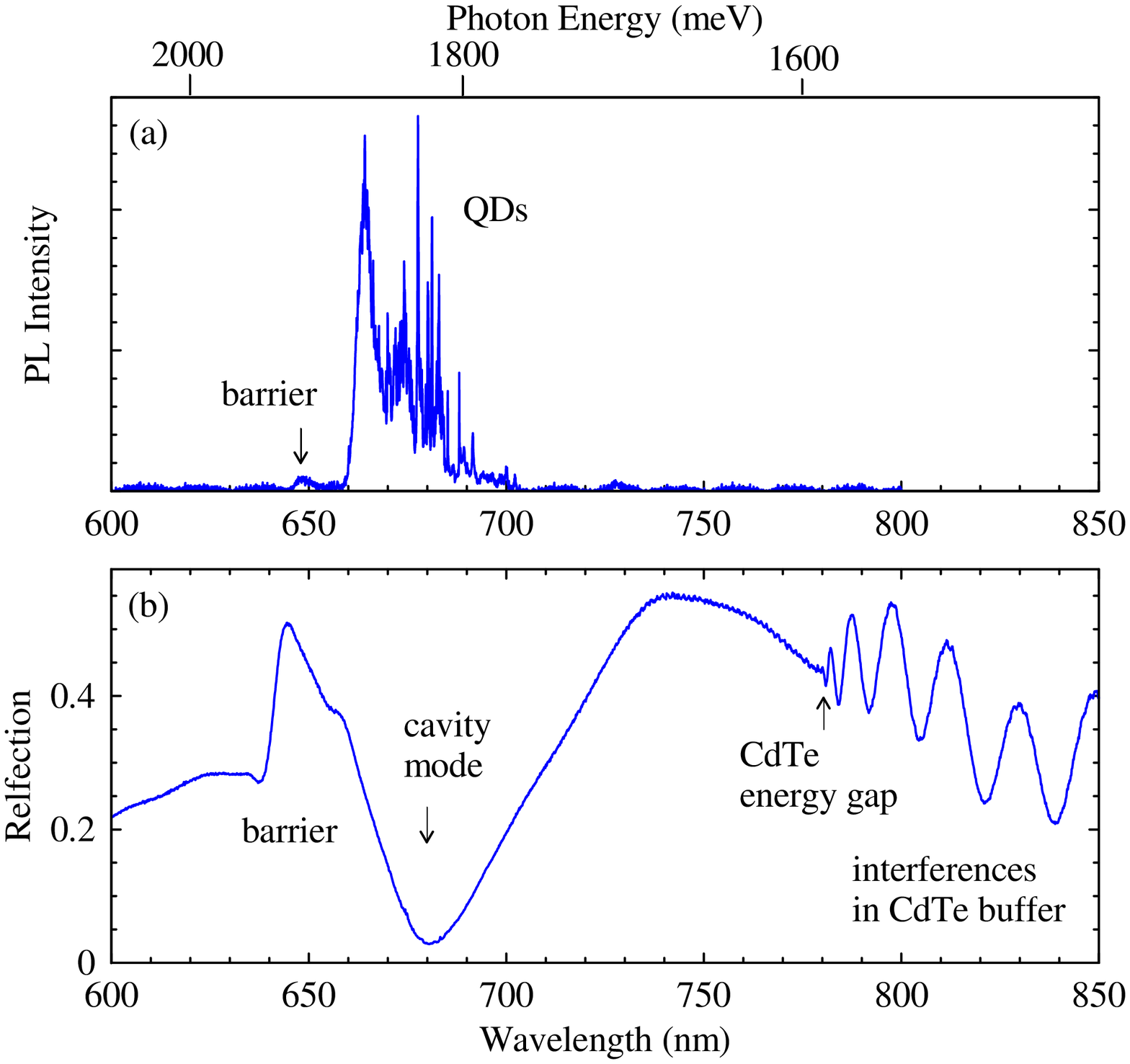}
\caption{(a) PL and (b) reflectance spectra measured at $T=7\,$K.
PL shows a multitude of sharp lines corresponding to excitonic
transitions in individual QDs. QDs are in resonance with cavity mode
observed in reflectance. Reflectance spectrum displays a broad
(about 40\,nm wide) and deep (below 3\% of reflectance) cavity mode,
showing that the DBR acts as an AR photonic structure for spectral
region of QDs emission.} \label{fig:PL-ref}
\end{figure}

\paragraph{Photoluminescence and reflectance.}
Fig.\,\ref{fig:PL-ref} shows low temperature photoluminescence (PL)
and reflectance spectra of studied sample, both measured using
experimental setup equipped in a microscope objective with spatial
resolution of about 1$\mu$m\,\cite{Sawicki_APL_2015}. PL was excited
nonresonantly at 532\,nm using CW power of about 1\,$\mu$W, well
below the QD saturation. The reflectance was measured using a white
(halogen) lamp.

\begin{figure}[t]
 \centering
\includegraphics[width=0.7\columnwidth]{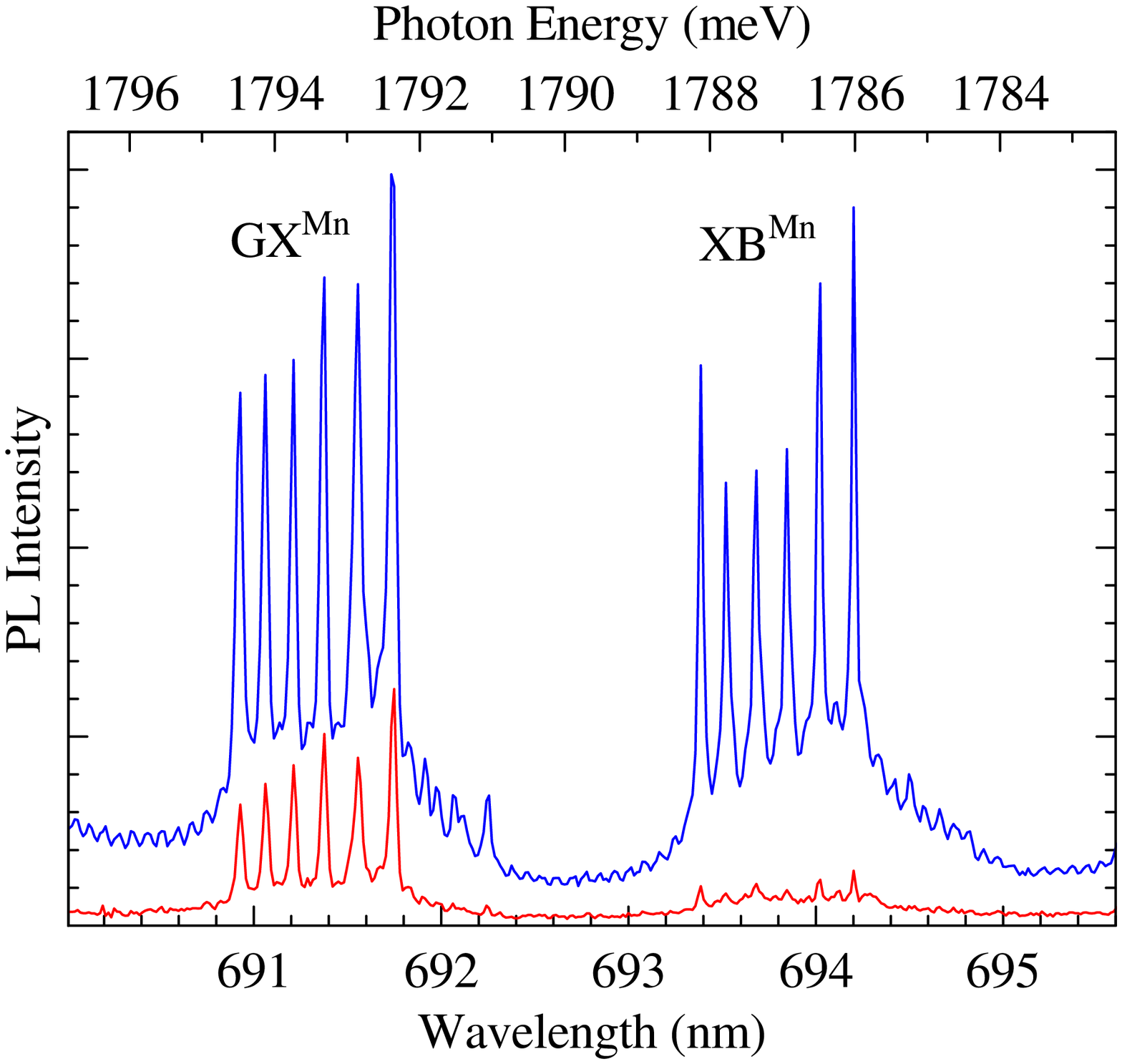}
\caption{Photoluminescence spectra of a single CdTe
/Cd$_{0.77}$Zn$_{0.13}$Mg$_{0.10}$Te QD with a single Mn ion,
measured for higher excitation power (upper spectrum) and lower
excitation power (lower spectrum). As expected, biexciton (XB$^{\rm
Mn}$) transitions exhibit higher sensitivity on excitation power
than the exciton (GX$^{\rm Mn}$) ones. Both clusters are sixfold split due to
$s$,$p$-$d$ exchange interaction between individual Mn$^{2+}$ ion and
carriers confined in a QD. } \label{fig:Mn}
\end{figure}

The PL reveals a signal from a thin CdTe layer at 665\,nm and sharp
lines of individual CdTe QDs in range 670-700\,nm. In our sample,
CdTe QDs are placed in Cd$_{0.77}$Zn$_{0.13}$Mg$_{0.10}$Te barrier,
which is also visible as a weak peak close to 650\,nm in the PL
spectrum and as a spectral wiggle in $R(\lambda)$ between 635 and 650\,nm. We
therefore estimate the confinement energy of around 50\,meV for the
excitons localized in QD-like states. Reflectance spectrum at the
investigated area of the wafer shows also a broad (about 40\,nm
wide) and deep (below 3$\%$ of reflectance) cavity mode centered at
670\,nm, in resonance with the emission of QDs. The obtained broad
and deep cavity mode, mimics an AR coating, but also provides a
local excitation intensity enhancement $\propto Q$ for the spectral
and spatial location of QDs. During the growth $R$ was minimized for
730\,nm (Fig.\,\ref{fig:structure}\,(a)) for the center of the sample,
 but as shown in Fig \ref{fig:PL-ref}\,(b), owing
to the thickness gradients, other parts of the wafer display a
spectrally shifted cavity mode. This permits to adjust low $R$ for
the desired spectral range by varying the position on the sample.

The sample offers a possibility to investigate and control an
exciton coherence coupled, via $s$,$p$-$d$ exchange interaction,
with an individual spin of a magnetic
dopant\cite{Besombes_PRL2004,Kudelski_2007_PRL,Kobak_NComms2014,TSmolenskiNC2016,Lafuente_Sampietro_2016_PRB,Fainblat_NL_2016}.
This is possible thanks to Mn $\delta$-doping introduced within the
CdTe QD layer. {Via micro-photoluminescence
spectroscopy we can recognize QDs containing Mn ions, and in
particular distinguish those incorporating exactly one such dopant.
To be more specific, QDs without magnetic ions exhibit sharp lines
(typical full-width at half-maximum (FWHM) below 0.1 meV),
originating from the exciton recombination. QDs with more than two
magnetic ions exhibit spectrally broader lines (typically FWHM above 1 meV),
which significantly shift in a magnetic field.\cite{Wojnar_PRB_2007}
For QDs with exactly two Mn ions, each excitonic transition gets
split into a manifold of 36 lines, owing to 6*6 possible spin states
of two Mn ions.\cite{Besombes_2Mn} Finally, a spectral fingerprint
of a QD with a single Mn ion is very characteristic: exciton
transitions display 6-fold splitting, due to $s$,$p$-$d$ exchange
interaction with the Mn ion. The latter exhibits 6 possible spin
states, corresponding to spin projection $\pm 5/2$, $\pm 3/2$, and
$\pm 1/2$ on quantization axis of excitons (growth
axis).\cite{Besombes_PRL2004} In our sample, majority of QDs do not
contain Mn ions and thus show single lines. We have not observed QDs
with two or more Mn ions. However, there is a small fraction (well
below 1\%) of QDs, which exhibit the PL pattern characteristic for
QDs containing exactly one individual Mn ion.} A typical example of PL measured at such magnetic QD is shown in Fig.\,\ref{fig:Mn}: the excitonic spectral
clusters corresponding to recombination of exciton-to-ground (GX$^{\rm Mn}$)  and biexciton-to-exciton states XB$^{\rm Mn}$ do display
the characteristic 6-fold splitting \cite{Besombes_PRL2004}. Spectra are measured as
a function of the excitation power. As expected, the biexciton
transitions exhibits a much stronger intensity dependence, than the
exciton ones (see Fig. \ref{fig:Mn}). Observation of a pronounced
and spectrally sharp biexciton complex, with a binding energy of
around 7\,meV, indicates that QDs forming from fluctuations of thin
CdTe layer in Cd$_{0.77}$Zn$_{0.13}$Mg$_{0.10}$Te barrier are in
fact well defined zero dimensional objects, which efficiently
localize even four carriers (two holes and two electrons). Other
excitonic complexes, corresponding to a negatively and/or positively
charged QDs, doped with single Mn$^{2+}$ ions have also been
identified (not shown).

\begin{figure}[t]
 \centering
\includegraphics[width=1\columnwidth]{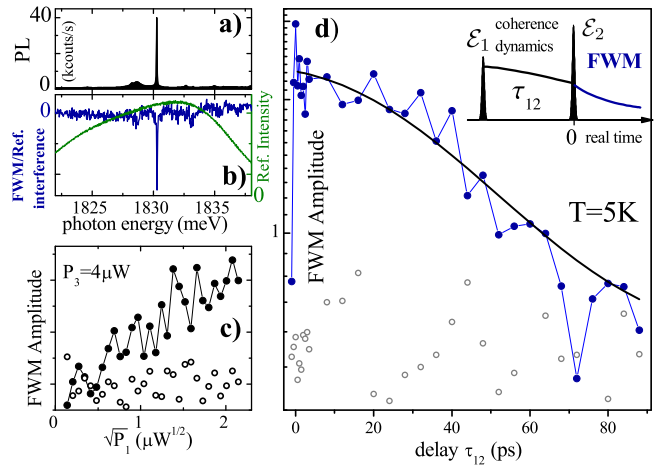}
\caption{Four-wave mixing of an individual CdTe QD (withut Mn ion)
in Cd$_{0.77}$Zn$_{0.13}$Mg$_{0.10}$Te barrier. (a) PL from a single
QD, showing a high emission flux of 40\,kcounts/s, indicating an
increased extraction efficiency. (b) FWM spectral interference
(blue) of the same transition as in (a) driven with the $\Ea$ and
$\Eb$ intensities $(P_1,\,P_2)=(2,\,6)\,\mu$W, $\tau_{12}=1.2\,$ps.
The $\Er$ spectrum is given with a green trace. (c) Time-integrated
FWM amplitude as a function of $\Ea$ amplitude $(\sqrt{P_1})$. (d)
Coherence dynamics of the transition shown in (b) measured with FWM
at $T=5\,$K: the applied pulse sequence is depicted in the inset.
The $\tau_{12}$-dependence of time-integrated FWM amplitude is shown
in blue, the noise level is given with open circles. We estimate the
dephasing time $T_2 =(210\,\pm\,40)\,$ps \label{fig:FWM}.}
\end{figure}

\paragraph{Four-wave mixing.}
To retrieve the coherent nonlinear response of individual QDs, we
employed the heterodyne spectral interferometry
\cite{Langbein_OL_2006}. Technical details regarding the current
implementation of the setup are provided in
Ref.\,[\onlinecite{FrasNatPhot16}]. To resonantly excite the QDs at
680-690\,nm with 0.3\,ps pulses, we use an optical parametric
oscillator (OPO model Inspire provided by Radiantis), pumped with a
Ti:Sapphire laser (model Tsunami Femto provided by Spectra-Physics).
The excitation spectrum generated by the OPO in shown in
Fig.\,\ref{fig:FWM}\,b (green). We form a pair of pulse trains $\Ea$
and $\Eb$ with a variable delay $\tau_{12}$ controlled with a
mechanical stage providing $\tau_{12}$ up to 1\,ns. $\Ea$ and $\Eb$
are modulated using acousto-optic deflectors, optimized for a VIS
range and driven at $\Omega_1=80\,$MHz and $\Omega_2=80.77\,$MHz,
respectively. Both beams are recombined into the same spatial mode
and are focussed to the diffraction limit at the sample surface
using a microscope objective (NA=0.65) installed on a XYZ
piezo-stage. We interfere the reflectance with the frequency shifted
reference beam $\Er$. In particular, by selecting the heterodyne
mixing at $2\Omega_2-\Omega_1=81.54$\,MHz, we select the beating
frequency corresponding to the FWM field, proportional to
$\Ea^{\star}\Eb\Eb$ in the lowest order. The FWM-$\Er$ interference is spectrally
resolved and detected by a CCD camera.

As candidates for FWM measurements, we pick up the transitions as
the one shown in Fig.\,\ref{fig:FWM}\,a, which feature an enhanced
flux of the PL, thus indicating an increased extraction efficiency
$(\eta\geq5)$ with respect to other QDs. By monitoring the
reflectance, we assure that the targeted transition lies at the
minimum of the cavity mode. As a result, we expect the weak FWM
field to be enhanced by a factor $\eta^{1/2}Q^{3/2}\simeq200$ with
respect to bare QDs, and thus capable to be induced with a
respectively lowered resonant intensity of \Ea and \Eb.

A representative example of the FWM interference at
$\tau_{12}=+1.2\,$ps is presented in Fig.\,\ref{fig:FWM}, with a
signal-to-noise of 10 after 300\,s integration. This is comparable
with previous FWM measurements on single QDs employing AR structures
\cite{Patton_PRB_2006, KasprzakNJP13}. In Fig.\,\ref{fig:FWM}\,c we
present FWM amplitude, as a function of $\Ea$ amplitude,
proportional to its area $\theta_1$. With increasing the latter, the
FWM is expected to yield Rabi oscillations \cite{FrasNatPhot16},
with the first maximum (minimum) at $\theta_1=\pi/2$ $(\pi)$. In our
case however, the accessible range of intensities is not sufficient
to attain significant $\theta_1$, so as to approach $\pi/2$ area.
The flopping is not observed and we remain in the $\chi^{(3)}$
driving regime of the FWM.

In the inset of Fig.\,\ref{fig:FWM}\,d we illustrate the two-pulse
sequence performed to infer the coherence dynamics of the same
transition ($\Er$ is set fixed, 1\,ps prior to $\Eb$): $\Ea$ induces
the coherence, which is converted into the density grating by $\Eb$.
The latter is then self-diffracted into the FWM polarization. By
measuring FWM as a function of $\tau_{12}$ one can access the
homogeneous broadening $\gamma=2\hbar/T_2$ (where $T_2$ denotes
dephasing time) within the inhomogeneously broadened distribution
via inferring photon echoes, which in case of a single QD are
created by a random spectral wandering \cite{Patton_PRB_2006,
KasprzakNJP13}. The data are fit by the model including both
homogeneous and inhomogeneous contributions to the spectral shape
\cite{KasprzakNJP13, MermillodPRL16}. The result, shown with a black
trace, yields the dephasing time of $T_2=(210\,\pm\,40)\,$ps and inhomogeneous broadening of $(50\,\pm\,20)\,\mu$eV (FWHM). A
large uncertainty of $T_2$ stems from a limited range of the
accessible delays, restricted by the time-resolution of the
spectrometer - the issue which could be overcome by introducing the
delay between $\Er$ and $\Eb$.\cite{JakubczykACSPhot16} Note the
decrease of the FWM during initial several ps, which is due to
dephasing with acoustic phonons.\cite{JakubczykACSPhot16} We observe
no FWM signal for $\tau_{12}<0$, which could rise from a two-photon
coherence\,\cite{MermillodPRL16} between the ground state and a
two-exciton (biexciton) state in a QD. The missing signal for $\tau_
{12}<0$, indicates that the QD is charged, allowing only for the
trion transition \cite{FrasNatPhot16}. Finally, it is worth to point
out an order of magnitude longer $T_2$, with respect to the
previous, and until now sole report available for the CdTe-platform
\cite{Patton_PRB_2006}.

\section{Discussion and conclusions}

Our structure represents the third successful approach to the growth
of photonic structures with QDs containing single Mn ions. The first
report has been presented in Ref.\,[\onlinecite{Pacuski_CGD2014}], where a
three-step process, involving two remote MBE facilities, was
established to grow ZnTe/MgTe/MgSe based microcavities with
CdTe/ZnTe QDs containing single Mn ions. In that case, growth
interruptions were due to missing molecular sources in growth
chambers: the first growth chamber was used to realize DBRs, the
second one was employed for the growth of Mn-doped QDs. Recently,
the same material system was used for the growth of Mn-doped QDs on
the top of 10 DBR pairs. The process was carried out using a single
MBE machine, but growth interruptions were anyway required to clean
the chamber from the residual Se atoms, which incorporated into QDs and impaired their optical properties \cite{Rousset_JAP_2016}.

Comparing to previous works, here the procedure is significantly
simplified, since Se is not involved in the growth of our
microcavity based on (Cd,Zn,Mg)Te. Therefore, the whole process is
realized in a single growth chamber, without growth interruptions.
It can be reproduced in any MBE growth chamber devoted to Te
compounds. One more important advantage of the present system is that
CdTe /Cd$_{0.77}$Zn$_{0.13}$Mg$_{0.10}$Te QDs with single Mn ion
presented here could be brought toward the spectral range of
Al$_2$O$_3$:Ti tunable femtosecond laser, i.e. above 700\,nm. This
could be particularly handy as it eliminates the requirement of
using OPO sources to perform resonant, ultrafast spectroscopy of
this material system. Finally, the design of antireflective photonic
structure presented in this work will be necessary for achieving a
satisfactory retrieval sensitivity of coherent nonlinear responses of magnetic QDs, where the exciton oscillator strength is distributed over a set of spin-assisted transitions.

\section{Acknowledgments}

%\footnotesize
This work was partially supported by the ministry of science and higher education under grants
Iuventus Plus  IP2014 040473 and IP2014 034573, and by the Polish National Science Centre under
decisions DEC-2015/18/E/ST3/00559, DEC-2011/02/A/ST3/00131, DEC-2014/13/N/ST3/03763,
DEC-2013/09/B/ST3/02603, DEC-2015/16/T/ST3/\-00506, UMO-2012/05/B/ST7/02155. The research was carried out with the use of CePT,
CeZaMat and NLTK infrastructures financed by the European Union - the European Regional
Development Fund within the Operational Programme "Innovative economy". VD, TJ and JK acknowledge
the financial support by the European Research Council (ERC) Starting Grant PICSEN (grant no. 306387).

%\vspace{1cm}

%\bibliographystyle{achemso}
%\bibliography{Bibliography_antireflective}

\providecommand{\latin}[1]{#1}
\providecommand*\mcitethebibliography{\thebibliography}
\csname @ifundefined\endcsname{endmcitethebibliography}
  {\let\endmcitethebibliography\endthebibliography}{}

\end{document}